\documentclass[aps,twocolumn]{revtex4}
\usepackage{amsmath,amssymb}
\begin{document}

\title{Charged rotating black holes in dilaton gravity}

\author{A. Sheykhi $^{1,2}$ \footnote{
sheykhi@mail.uk.ac.ir} and  N. Riazi $^{1}$ \footnote{
riazi@physics.susc.ac.ir}}
\address{$^1$ Physics Department and Biruni Observatory, Shiraz University, Shiraz 71454, Iran\\
         $^2$ Department of Physics, Shahid Bahonar University, Kerman, Iran}
\begin{abstract}
We consider charged black holes with \emph{curved horizons}, in
five dimensional dilaton gravity in the presence of Liouville-type
potential for the dilaton field. We show how, by solving a pair of
coupled differential equations, infinitesimally small angular
momentum can be added to these static solutions to obtain
charged rotating dilaton black hole solutions. In the absence of
dilaton field, the non-rotating version of the solution reduces to
the five dimensional Reissner-Nordstr\"{o}m black hole, and the
rotating version reproduces the five dimensional Kerr-Newman
modification thereof for small rotation parameter. We also compute
the angular momentum and the angular velocity of these rotating
black holes which appear at the first order.
\end{abstract}

\maketitle 
\section{Introduction}
Inspired by the string theory that gravity is not given by the
Einstein action, at least at sufficiently high energies, a lot of
investigations have been done in the literature in recent years.
The low-energy limit of the string theory leads to the Einstein
gravity, coupled nonminimally to a scalar dilaton field
\cite{Wit1}. When a dilaton is coupled to Einstein-Maxwell theory,
it has profound consequences for the black hole solutions. This
fact may be seen in the case of rotating Einstein-Maxwell-dilaton
(EMD) black holes of Kaluza-Klein theory with coupling constant
$\alpha =\sqrt{3}$ which does not possess the gyromagnetic ratio
$g=2$ of Kerr-Newman black hole \cite{Gib,Fr,Hor1}. Thus it is
worth finding black hole solutions of EMD gravity for an arbitrary
value of dilaton coupling constant and investigate how the
properties of black holes are modified when a dilaton is present.

Exact charged dilaton black hole solutions in the absence of
dilaton potential have been constructed by many authors
\cite{Gib2,koi,Bri,Gar, Gre, Rak,Bou}. The dilaton changes the
causal structure of the black hole and leads to curvature
singularities at finite radii. These black holes are
asymptotically flat. In recent years, non-asymptotically flat
black hole spacetimes are attracting much interest. A motivation
to investigate non-asymptotically flat, nonasymptotically AdS
solutions of Einstein gravity is that these might lead to possible
extensions of AdS/CFT correspondence. Indeed, it has been
speculated that the linear dilaton spacetimes, which arise as
near-horizon limits of dilatonic black holes, might exhibit
holography \cite{Ah}. Another motivation is that such solutions
may be used to extend the range of validity of methods and tools
originally developed for, and tested in the case of,
asymptotically flat or asymptotically AdS black holes. Black hole
spacetimes which are neither asymptotically flat nor (anti)- de
Sitter (A)dS have been found and investigated by many authors. The
uncharged solutions have been found in \cite{MW,MW2,MW3}, while
the charged solutions have been considered in \cite{PW,PW2,PW3}.
In the presence of Liouville-type potential, static charged
solutions of EMD gravity have been discovered with positive, zero
or negative constant curvature horizons
\cite{CHM,Yaz,Yaz2,Shey1,Shey2,Cai,Cai2,Cai3}. Recently, the
properties of these black hole solutions which are neither
asymptotically flat nor (A)dS have been disclosed in
\cite{Clem,Clem2}.

These exact solutions \cite{Gib2,koi,Bri,Gar,Gre,Rak,Bou} and
\cite{MW,MW2,MW3,PW,PW2,PW3,CHM,Yaz,Yaz2,Cai,Cai2,Cai3,Clem,Clem2}
are all static. Charged rotating dilaton black holes with
\emph{curved horizon} have not been constructed in four or higher
dimensions for an arbitrary coupling constant and arbitrary
rotation parameter. Indeed, exact magnetic rotating solutions have
been considered in three dimensions \cite{Dia}, while exact
rotating solutions of EMD gravity have been obtained only for some
limited values of the coupling constant
\cite{Fr}\cite{Har,Har1,Har2,Har3}. For general dilaton coupling
constant, the properties of charged dilaton black holes only in
four dimensions with small angular momentum
\cite{Hor1,Shi,Shi1,SR} or small charge \cite{Cas} have been
investigated. When the horizons are flat, magnetic and electric
rotating solutions in four-dimensional EMD gravity have also been
constructed in \cite{Deh1} and \cite{Deh2}, respectively.
Recently, this solutions have been generalized to the
$(n+1)$-dimensional EMD gravity \cite{SDRP,SDR}. These solutions
(\cite{SDRP,SDR}) are not black holes and describe charged
rotating \emph{black branes} with \emph{flat horizons}. Until now,
charged rotating dilaton black holes solutions with \emph{curved
horizons} for an arbitrary value of dilaton coupling constant in
more than four dimensions have not been constructed. The
motivation for studying higher dimensional black holes comes from
developments in string/M-theory, which is believed to be the most
consistent approach to quantum theory of gravity in higher
dimensions. It was argued that black holes may play a crucial role
in the analysis of dynamics in higher dimensions as well as in the
compactification mechanisms. In particular, to test novel
predictions of string/M-theory microscopic black holes may serve
as good theoretical laboratories. It has been thought that the
statistical-mechanical calculation of the Bekenstein-Hawking
entropy for a class of supersymmetric black holes in five
dimensions is one of the remarkable results in string theory
\cite{Strom,Strom1}. Another motivation on studying higher
dimensional black holes originates from the braneworld scenarios,
as a new fundamental scale of quantum gravity. An interesting
consequence of these models is the possibility of mini black hole
production at future colliders \cite{Dim}. These serve as our main
motivation to explore of the effects of dilaton field on the
properties of charged rotating black holes in higher dimension. In
this regard, as a new step to shed some light on this issue for
further investigation, we report a new class of solution of the
Einstein-Maxwell gravity coupled to a dilaton field which
describes an electrically charged, slowly rotating black hole with
\emph{curved horizon} in five dimensions with arbitrary value of
coupling constant $\alpha$. We shall also investigate the effects
of dilaton field as well as rotation parameter on the physical
quantities such as temperature, entropy, angular momentum and the
angular velocity of these rotating black holes.


\section{Field equations and solutions}

Our starting point is the following action
\begin{eqnarray}
S &=&-\frac{1}{16\pi }\int_{\mathcal{M}}d^{n}x\sqrt{-g}\left( \mathcal{%
R}\text{ }-\frac{4}{n-2}(\nabla \Phi )^{2} \right. \nonumber
\\
&& \left. -V(\Phi )-e^{-4\alpha \Phi
/(n-2)}F_{\mu \nu }F^{\mu \nu }\right)   \nonumber \\
&&-\frac{1}{8\pi }\int_{\partial \mathcal{M}}d^{n-1}x\sqrt{-\gamma
}\Theta (\gamma ),  \label{act1}
\end{eqnarray}
where ${\cal R}$ is the Ricci scalar curvature, $\Phi$ is the
dilaton field and $V(\Phi)$ is a potential for $\Phi$. $\alpha $
is a constant determining the strength of coupling of the scalar
and electromagnetic field, $F_{\mu \nu }=\partial _{\mu }A_{\nu
}-\partial _{\nu }A_{\mu }$ is the electromagnetic  field tensor
and $A_{\mu }$ is the electromagnetic potential. The last term in
Eq. (\ref{act1}) is the Gibbons-Hawking boundary term which is
chosen such that the variational principle is well-defined. The manifold $%
\mathcal{M}$ has metric $g_{\mu \nu }$ and covariant derivative
$\nabla _{\mu }$. $\Theta $ is the trace of the extrinsic
curvature $\Theta ^{ab}$ of any boundary(ies) $\partial
\mathcal{M}$ of the manifold $\mathcal{M}$, with induced metric(s)
$\gamma _{ab}$. In this paper, we consider the action (\ref{act1})
with a Liouville type potential,
\begin{equation}
V(\Phi )=2\Lambda e^{2\beta \Phi},  \label{v1}
\end{equation}
where $\Lambda $ and $\beta$ are arbitrary constants. One may
refer to $\Lambda$ as the cosmological constant, since in the
absence of the dilaton field ($\Phi =0$) the action (\ref{act1})
reduces to the action of Einstein-Maxwell gravity with
cosmological constant. The equations of motion can be obtained by
varying the action (\ref{act1}) with respect to the gravitational
field $g_{\mu \nu }$, the dilaton field $\Phi $ and the gauge
field $A_{\mu }$ which yields the following field equations
\begin{eqnarray}\label{FE1}
{\cal R}_{\mu\nu}=
\frac{4}{n-2}\left(\partial_{\mu}\Phi\partial_{\nu}\Phi
+\frac{1}{4}g_{\mu\nu}V(\Phi)\right) \nonumber \\
+2e^{-4\alpha\Phi/(n-2)}\left( F_{\mu\eta} F_{\nu}^{\text{ } \eta}
-\frac{g_{\mu\nu}}{2(n-2)}F_{\lambda\eta}F^{\lambda\eta}\right),
\end{eqnarray}
\begin{equation}\label{FE2}
\nabla ^{2}\Phi =\frac{n-2}{8}\frac{\partial V}{\partial
\Phi}-\frac{\alpha}{2}e^{-4\alpha\Phi/(n-2)}F_{\lambda\eta}F^{\lambda\eta},
\end{equation}
\begin{equation}\label{FE3}
\nabla_{\mu}\left(e^{-4\alpha\Phi/(n-2)} F^{\mu\nu}\right)=0.
\end{equation}
We wish to find five dimensional rotating solutions of the above
field equations, thus we set $n=5$. To this end,  we first study
the non-rotating black hole for arbitrary $\alpha$, then, we
consider the effect of adding a small amount of rotation parameter
$a$ to the black hole. We will discard any terms involving $a^2$
or higher power in $a$. For infinitesimal rotation, we can solve
Eqs. (\ref{FE1})-(\ref{FE3}) to first order in the angular
momentum parameter $a$. This is because most of the metric
components depend only on $a^2$. In fact, many of the interesting
physical quantities also depend only on $a^2$, however we can
still extract some useful information from the first-order
solutions. Inspection of the five dimensional Kerr-Newman
solutions shows that the only term in the metric changes to $O(a)$
is $g_{t\phi}$. Similarly, the dilaton field does not change to
$O(a)$ and $A_{\phi}$ is the only component of the vector
potential that changes. Therefore, for infinitesimal angular
momentum up to $O(a)$, we can take the following form of the
metric
\begin{eqnarray}\label{metric}
ds^2 &=&-U(r)dt^2+{dr^2\over U(r)}- 2 a f(r)\sin^{2}{\theta}dt
d{\phi}\nonumber \\
 &&+ R^2(r)\left(d\theta^2 + \sin^2\theta d\phi^2 +cos^2\theta
d\psi^2\right).
\end{eqnarray}
The unknown functions $U(r)$, $R(r)$ and $f(r)$ should be
determined. In the particular case $a=0$, this metric reduces to
the static and spherically symmetric cases. For small $a$, we can
expect to have solutions with $U(r)$ still a function of $r$
alone.

The $t$ component of the Maxwell equations can be integrated
immediately to give
\begin{equation}\label{Ftr}
 F_{tr}= \frac{q e^{4\alpha\Phi/3}}{ R^3(r)},
\end{equation}
where $q$ is  an integration constant related to the electric
charge of the solutions. Defining the electric charge via
\begin{equation}\label{Q}
Q = \frac{1}{4\pi}\int_{s^3} e^ {-4\alpha \Phi/3}  \text{}^{*} F
d{\Omega},
\end{equation}
where $*$ is the Hodge dual and $ s^{3}$ is any 3-sphere defined
at spatial infinity, with its volume element denoted by
$d{\Omega}$. Then the electric charge of the black hole will be
\begin{equation}\label{q}
Q = \frac{q w_3}{4\pi},
\end{equation}
where $w_{3}$ represents the volume of the unit 3-sphere. In
general, in the presence of rotation, there is also a vector
potential in the form
\begin{equation}\label{Aphi}
 A_{\phi}=a q h(r)\sin^2\theta.
\end{equation}
It is worth noting that for infinitesimal rotation parameter, the
electric field (\ref{Ftr}) does not change from the static case.

With the metric (\ref{metric}) and the Maxwell fields (\ref{Ftr})
and (\ref{Aphi}), the field equations (\ref{FE1})-(\ref{FE3})
reduce to the following system of coupled ordinary differential
equations
\begin{eqnarray}\label{ODE1}
&&R^{6}\frac{d^2U}{dr^2}+R^{5}{\frac{dR}{dr}}{\frac{dU}{dr}}-4
R^{4}U ({\frac{dR}{dr}})^{2} \nonumber \\
&&-2 R^{5}U ({\frac{d^{2}R}{dr^{2}}}) =-4{R^4} + 4q^{2}
e^{{4\alpha\Phi}/{3}},
\end{eqnarray}
\begin{equation}\label{ODE2}
{1\over R^3}{d\over dr }\left(U {dR^3\over dr } \right)=
\frac{6}{R^2}-V(\Phi)- 2 e^{4\alpha\Phi/3}\frac{q^2}{R^6},
\end{equation}
\begin{equation}\label{ODE3}
\frac{1}{R^3}{d\over dr }\left(R^3 U {d\Phi\over dr } \right)
=\frac{3}{8}\frac{d{V}}{d\Phi}+\alpha
e^{{4\alpha\Phi}/{3}}\frac{q^2}{R^6},
\end{equation}
\begin{equation}\label{ODE4}
{1\over R}{d^2R\over dr^2} =-\frac{4}{9} \left({d\Phi\over dr
}\right)^2.
\end{equation}
In addition, we have two coupled differential equations for
functions $f(r)$ and $h(r)$.
\begin{eqnarray}\label{ODE5}
&&R^{3}\frac{d^2f}{dr^2}+R^{2}{\frac{dR}{dr}}{\frac{df}{dr}}-4fR({\frac{dR}{dr}})^{2}\nonumber
\\&& -2fR^{2}\frac{d^2R}{dr^2}
-4q^{2}{\frac{dh}{dr}}=0,
\end{eqnarray}
\begin{eqnarray}\label{ODE6}
&& R^{2}\frac{d}{dr}(Ue^{{-4\alpha\Phi}/{3}}\frac{dh}{dr})-
R\frac{d}{dr}(\frac{f}{R^2})\nonumber \\
&& +\frac{1}{2}\left(U \frac{dh}{dr} \frac{dR^2}{dr}
-8h\right)e^{{-4\alpha\Phi}/{3}}=0.
\end{eqnarray}
These two equations which arise from the presence of $A_\phi$,
appear only when $a\neq 0$, while the other equations were there
also in the static, spherically symmetric case.


In order to solve eqs. (\ref{ODE1})-(\ref{ODE4}), following
\cite{CHM}, we make the ansatz
\begin{equation}\label{R1}
R(r)=r^{N},
\end{equation}
where $N$ is a constant. Using (\ref{R1}), one can easily show
that  Eqs. (\ref{ODE1})-(\ref{ODE4}), have solutions of the form
\cite{CHM}
\begin{eqnarray}\label{U2}
U(r)&=&
r^{\frac{2\alpha^2}{1+\alpha^2}}\left(\frac{2(1+\alpha^2)^2}{(2+\alpha^2)(1-\alpha^2)}
-\frac{4M(1+\alpha^2)}{3}r^{\frac{-(\alpha^2+2)}{1+\alpha^2}}\right.
\nonumber
\\
&&\left.+ \frac{2q^2(1+\alpha^2)^2e^{4\alpha b/3}}{3(2+\alpha^2)}
r^{\frac{-2(\alpha^2+2)}{\alpha^2+1}} \right),
\end{eqnarray}
\begin{eqnarray}
\Phi(r)&=&b-\frac{3\alpha}{2(1+\alpha^2)}\ln(r),\\
R(r)&=& r^{1/(\alpha^2+1)}.
\end{eqnarray}
with $b$ a constant and $M$ is the quasi-local mass of the black
hole \cite{CHM,BY}. In order to fully satisfy the system of
equations, we must have $\beta={2}/{3\alpha}$. The constant $b$ is
related to the $\Lambda$ parameter via
\begin{equation}
\Lambda=\frac{3\alpha^2 e^{-4b /3\alpha}}{\alpha^2-1}.
\end{equation}
It is worth noting that the solution does not exist for string
case where $\alpha=1$. The $ \alpha \rightarrow \infty$ limit
produces (anti)-de Sitter behavior for zero $M$ and $q$. On the
other hand, in the absence of dilaton field $(\alpha=0)$, the
solution becomes
\begin{eqnarray}\label{alpha0}
U(r)&=& 1-\frac{4M}{3r^2}+\frac{q^2}{3r^4},\nonumber\\
R(r)&=& r,
\end{eqnarray}
which is the the five dimensional Reissner-Nordstr\"{o}m black
hole solutions for vanishing rotation parameter $a$.  Horizons are
located at
\begin{eqnarray}\label{horizon1}
&& r_{h}^{\frac{\alpha^2+2}{\alpha^2+1}}=\frac{(1-\alpha^2)(2+\alpha^2)M}{3(1+\alpha^2)}\nonumber\\
&&\times\left(1\pm \sqrt{1-\frac{3q^2
e^{\frac{4b\alpha}{3}}(1+\alpha^2)^2}{M^2(1-\alpha^2)(2+\alpha^2)^2}}\right).
\end{eqnarray}
If $\alpha^2<1$, then there are two horizons, while we have a
single horizon for $\alpha^2>1$. There is  an extremal limit for
the electric charge
\begin{equation}\label{qextr}
q_{ext}^2=\frac{(1-\alpha^2)(2+\alpha^2)^2e^{\frac{-4b\alpha}{3}}}{3(1+\alpha^2)^2}M_{ext}^2.
\end{equation}
In the case $q^2>q_{ext}^2$, we have naked singularity. The metric
corresponding to (\ref{U2}) is neither asymptotically flat nor
(anti)-de Sitter. In order to study the general structure of these
solutions, we first look for the curvature singularities in the
presence of dilaton gravity up to $O(a)$. It is easy to show that
the Kretschmann scalar $R_{\mu \nu \lambda \kappa }R^{\mu \nu
\lambda \kappa }$ diverges at $r=0$, it is finite for $r\neq 0$
and goes to zero as $r\rightarrow \infty $. Also, it is notable to
mention that the Ricci scaler is finite every where except at
$r=0$, and goes to zero as $r\rightarrow \infty $. Therefore
$r=r_{h}$ is a regular horizon and we have an essential
singularity located at $r=0$. Note that the dilaton field is
regular on the horizons, too.

Black hole entropy typically satisfies the so called area law of
the entropy \cite{Beck,Beck1,Beck2}, which states that the entropy
is a quarter of the event horizon area. This near universal law
applies to almost all kinds of black holes and black holes in
Einstein gravity \cite{hunt,hunt1,hunt2,hunt3}. Since the surface
gravity and area of the event horizon do not change to $O(a)$, one
can easily show that the temperature and the entropy of black hole
on the outer event horizon can be written as
\begin{eqnarray}
T &=& {1\over 4\pi} {dU \over
dr}(r_{h})=\frac{(2+\alpha^2)M}{3\pi}
r_{h}^{-\frac{5+\alpha^2}{1+\alpha^2}}\nonumber\\
&&\times\left(r_{h}^{\frac{2+\alpha^2}{1+\alpha^2}}-\frac{(1+\alpha^2)
q^2e^{\frac{4b\alpha}{3}}}{(2+\alpha^2)M}\right), \\
S &=& \frac{\pi^2}{2}{r_{h}^{3/(\alpha^2+1)}}.\label{TS2}
\end{eqnarray}
Note that the temperature vanishes in the extremal limit when
$q=q_{ext}$.

Here, we are interested in finding the rotating version of this
static solution, that is to say, in solving the corresponding
coupled equations for two unknown functions $f(r)$ and $h(r)$. For
arbitrary value of the dilaton coupling constant $\alpha$, we
could obtain the following exact solution
\begin{eqnarray}\label{fhV}
f(r)&=&\frac{4M(\alpha^2+2)e^{\frac{-4b\alpha}{3}}}{3}r^{\frac{\alpha^2-2}{\alpha^2+1}}
-\frac{2q^2(\alpha^2+1)}{3}r^{-\frac{4}{\alpha^2+1}},\nonumber\\
h(r)&=& r^{-\frac{\alpha^2+2}{\alpha^2+1}}.
\end{eqnarray}
Note that for $\alpha^2<2$, this solution decreases with increase
of $r$. For ($\alpha=0$), the non-rotating version of the solution
reduces to the five dimensional Reissner-Nordstr\"{o}m  black hole
(\ref{alpha0}), and the rotating version (\ref{fhV}) reproduces
the five dimensional Kerr-Newman modification thereof for small
rotation parameter $a$ \cite{Aliev1} (see also \cite{Aliev2})
\begin{eqnarray}\label{KN}
f(r)&=&\frac{8M}{3r^2}-\frac{2q^2}{3r^4},\nonumber\\
h(r)&=&\frac{1}{r^2}.
\end{eqnarray}

Finally, we study the physical properties of these solutions, by
computing the angular velocity of the solutions at the horizons
and the value of the angular momentum in the general situation
when $\alpha\neq0$. The angular velocity at the horizon $r=r_{h}$
is given in the leading order by
\begin{eqnarray}
\Omega_{h}&=&\frac{g_{t\phi}(r=r_{h},\theta=\pi/2)}{R^2(r_{h})}\nonumber\\
&=&\left(-\frac{4Ma(\alpha^2+2)}{3}e^{\frac{-4b\alpha}{3}}r_{h}^{\frac{\alpha^2-4}{\alpha^2+1}}\right.
\nonumber
\\
&& \left.
+\frac{2q^2a(\alpha^2+1)}{3}r_{h}^{-\frac{6}{\alpha^2+1}}\right).
\end{eqnarray}
while the angular momentum of the black hole can be calculated
through the use of the quasi-local formalism of the Brown and York
\cite{BY}. According to the quasilocal formalism, the quantities can be
constructed from the information that exists on the boundary of a
gravitating system alone. Such quasilocal quantities will
represent information about the spacetime contained within the
system boundary, just like the Gauss's law. In our case the finite
stress-energy tensor can be written as
\begin{equation}
T^{ab}=\frac{1}{8\pi }\left(\Theta^{ab}-\Theta\gamma ^{ab}\right)
, \label{Stres}
\end{equation}
which is obtained by variation of the action (\ref{act1}) with respect
to the boundary metric $\gamma _{ab}$. To compute the
angular momentum of the spacetime, one should choose a spacelike surface $%
\mathcal{B}$ in $\partial \mathcal{M}$ with metric $\sigma _{ij}$,
and write the boundary metric in ADM (Arnowitt-Deser-Misner)
form:
\[
\gamma _{ab}dx^{a}dx^{a}=-N^{2}dt^{2}+\sigma _{ij}\left( d\varphi
^{i}+V^{i}dt\right) \left( d\varphi ^{j}+V^{j}dt\right) ,
\]
where the coordinates $\varphi ^{i}$ are the angular variables
parameterizing the hypersurface of constant $r$ around the origin,
and $N$ and $V^{i}$ are the lapse and shift functions
respectively. When there is a Killing vector field $\mathcal{\xi
}$ on the boundary, then the quasilocal conserved quantities
associated with the stress tensors of Eq. (\ref{Stres}) can be
written as
\begin{equation}
Q(\mathcal{\xi )}=\int_{\mathcal{B}}d^{n-2}\varphi \sqrt{\sigma }T_{ab}n^{a}%
\mathcal{\xi }^{b},  \label{charge}
\end{equation}
where $\sigma $ is the determinant of the metric $\sigma _{ij}$, $\mathcal{%
\xi }$ and $n^{a}$ are the Killing vector field and the unit
normal vector on the boundary $\mathcal{B}$. For boundaries with
rotational ($\varsigma =\partial /\partial \varphi $) Killing
vector fields, one obtains the quasilocal angular momentum
\begin{eqnarray}
J &=&\int_{\mathcal{B}}d^{n-2}\varphi \sqrt{\sigma
}T_{ab}n^{a}\varsigma ^{b},  \label{Angtot}
\end{eqnarray}
provided the surface $\mathcal{B}$ contains the orbits of
$\varsigma $. Finally, the angular momentum of the black holes can
be calculated through the use of Eq. (\ref{Angtot}). We find
\begin{equation}
J =\frac{(\alpha^2+2)(4-\alpha^2)e^{-4b\alpha/3}w_{3}}{12\pi
(\alpha^2+1)} M a.  \label{J}
\end{equation}
For $a=0$, the angular momentum  vanishes, and therefore $a$ is
the rotational parameter of the black hole. It is worth noting
that $J\propto M a$, as one expected, and reduces to the case of
slowly rotating five dimensional Kerr solution in the absence of
dilaton field $(\alpha=0)$.

\section{Summary and Conclusion}
As noted in the introduction, exact, charged rotating dilaton black hole solutions with
\emph{curved horizons} for an arbitrary value of dilaton coupling
constant in more than four dimensions have not been constructed.
In this paper we studied charged black hole solutions in five
dimensional Einstein-Maxwell-dilaton gravity. These black holes
have unusual asymptotics. They are neither asymptotically flat nor
(anti-) de Sitter. Then, we considered the effect of adding a
small amount of rotation parameter $a$ to the black hole. We
discarded any terms involving $a^2$ or higher power in $a$.
Inspection of the Kerr-Newman solutions shows that the only term
in the metric changes to $O(a)$ is $g_{t\phi}$. Similarly, the
dilaton does not change to $O(a)$ and $A_{\phi}$ is the only
component of the vector potential that change to $O(a)$. For small
angular momentum, the field equations led to the coupled
differential equations (\ref{ODE5}) and (\ref{ODE6}) for two
unknown functions $f(r)$ and $h(r)$, for which we find a class of
solutions in the presence of Liouville-type potential. We showed
that in the absence of dilaton field $(\alpha=0)$, the
non-rotating version of the solution reduces to the five
dimensional Reissner-Nordstr\"{o}m black hole, and the rotating
version reproduces the five dimensional Kerr-Newman modification
for small $a$. We computed temperature and entropy of black hole,
which did not change to $O(a)$ from the static case. We also
obtained the angular momentum and the angular velocity of these
rotating black holes which appear at the first order of rotation
parameter $a$. It is notable to mention that the five dimensional
charged rotating dilaton black hole solutions obtained here have
small angular momentum. Thus, it would be interesting if one can
construct charged rotating dilaton black holes with arbitrary
rotation parameter. One can also attempt to extend these solutions
to higher dimensional ($n>5$) rotating dilaton black holes with
curved horizons.

\acknowledgments{This work was supported in part by Shiraz
University and also by Shahid Bahonar University.}


\end{document}